\documentclass[a4paper]{jpconf}
\usepackage{graphicx}
\usepackage{amssymb}
\usepackage{xcolor}
\usepackage{multirow}
\usepackage{siunitx}
\usepackage[numbers,square,sort&compress]{natbib}
\begin{document}

\title{Kinetic Inductance Detectors and readout electronics for the OLIMPO experiment}
\author{A Paiella$^{1,2}$, E S Battistelli$^{1,2}$, M G Castellano$^3$, I Colantoni$^{3,}$\footnote[6]{Present address: School of Cosmic Physics, Dublin Institute for Advanced Studies, 31 Fitzwilliam Place, D02 XF86, Dublin, Ireland.}, \\F Columbro$^{1,2}$, A Coppolecchia$^{1,2}$, G D'Alessandro$^{1,2}$, \\P de Bernardis$^{1,2}$, S Gordon$^4$, L Lamagna$^{1,2}$, H Mani$^4$, S Masi$^{1,2}$, \\P Mauskopf$\:^{4,5}$, G Pettinari$^3$, F Piacentini$^{1,2}$ and G Presta$^1$}
\address{$^1$ Dipartimento di Fisica, \emph{Sapienza} Universit\`a di Roma, P.le A. Moro 2, 00185 Roma, Italy}
\address{$^2$ Istituto Nazionale di Fisica Nucleare, Sezione di Roma, P.le A. Moro 2, 00185 Roma, Italy}
\address{$^3$ Istituto di Fotonica e Nanotecnologie - CNR, Via Cineto Romano 42, 00156 Roma, Italy}
\address{$^4$ School of Earth and Space Exploration, Arizona State University, Tempe, AZ 85287, USA}
\address{$^5$ Department of Physics, Arizona State University, Tempe, AZ 85257, USA}

\ead{alessandro.paiella@roma1.infn.it}

\begin{abstract}
Kinetic Inductance Detectors (KIDs) are superconductive low--temperature detectors useful for astrophysics and particle physics. We have developed arrays of lumped elements KIDs (LEKIDs) sensitive to microwave photons, optimized for the four horn--coupled focal planes of the OLIMPO balloon--borne telescope, working in the spectral bands centered at \SI{150}{GHz}, \SI{250}{GHz}, \SI{350}{GHz}, and \SI{460}{GHz}. This is aimed at measuring the spectrum of the Sunyaev--Zel'dovich effect for a number of galaxy clusters, and will validate LEKIDs technology in a space--like environment. Our detectors are optimized for an intermediate background level, due to the presence of residual atmosphere and room--temperature optical system and they operate at a temperature of \SI{0.3}{K}. The LEKID planar superconducting circuits are designed to resonate between 100 and \SI{600}{MHz}, and to match the impedance of the feeding waveguides; the measured quality factors of the resonators are in the $10^{4}-10^{5}$ range, and they have been tuned to obtain the needed dynamic range. The readout electronics is composed of a \emph{cold part}, which includes a low noise amplifier, a dc--block, coaxial cables, and
power attenuators; and a \emph{room--temperature part}, FPGA--based, including up and down-conversion microwave components (IQ modulator, IQ demodulator, amplifiers, bias tees, attenuators). In this contribution, we describe the optimization, fabrication, characterization and validation of the OLIMPO detector system.
\end{abstract}

\section{Introduction}

In the last thirty years, precision cosmology has achieved important goals through measurements of the Cosmic Microwave Background (CMB) radiation such as its spectrum \cite{0004-637X-473-2-576}, the anisotropies \cite{refId01}, the E--mode component of the polarization \cite{refId0}, and the B--mode component of the polarization due to gravitational lensing from dark matter structure \cite{0004-637X-833-2-228}. Yet, the B-mode power spectrum from inflation and the spectral distortions still remain elusive as well as the spectroscopic measurement of the Sunyaev--Zel'dovich (SZ) effect.

The OLIMPO experiment \cite{Coppolecchia2013} is aimed at measuring the SZ effect which is a CMB anisotropy in the direction of galaxy clusters, due to the inverse Compton scattering of low energy CMB photons by the high energy electrons of the hot gas present in the intra--cluster medium. SZ effect measurements represent an interesting tool to study the morphological and dynamical state of  clusters, to probe the CMB temperature evolution with the redshift, to constraint cosmological parameters, and to search for previously unknown clusters by looking at their SZ signature in the microwave sky \cite{1475-7516-2018-04-020, 1475-7516-2018-04-019}. 

OLIMPO measures SZ signals with a technique so far unattempted in this kind of obervations: it performs a spectroscopic map of the SZ effect with a differential interferometric instrument, working above the atmosphere, and provides efficient and unbiased decontamination of the SZ and CMB signals from all the foregrounds along the same line of sight \cite{deBernardis}, thus increasing the accuracy on the estimate of the astrophysical quantities involved in the physics of the effect.

The OLIMPO experiment has been, therefore, designed as a large balloon--borne mm--wave observatory, with a \SI{2.6}{m} aperture telescope, equipped with a room--temperature differential Fourier transform spectrometer (DFTS) \cite{schillaci2014}, and four low--temperature detector arrays, centered at 150, 250, 350, and \SI{460}{GHz}, exploring the negative, zero, and positive regions of the SZ spectrum. The detector arrays, consisting of horn--coupled lumped element kinetic inductance detectors (LEKIDs), are cooled to about \SI{300}{mK} by a $^{3}$He fridge, accomodated inside a wet N$_2$ plus $^{4}$He cryostat. The detector arrays are fed and read out coupled by means of two independent bias--readout lines and two FPGA--based electronics.  

Kinetic inductance detectors are superconductive photon detectors, where the radiation is detected by sensing changes of the kinetic inductance. A superconductor, cooled below its critical temperature $T_{c}$, presents two populations of electrons: quasiparticles and Cooper pairs, bound states of two electrons with binding energy $2\Delta_{0}=3.528\:k_{B}T_{c}$. If pair-breaking radiation ($h\nu>2\Delta_{0}$) is absorbed in a superconducting film, it breaks Cooper pairs, producing a change in the population relative densities, and thus in the kinetic inductance. For these reasons, in the lumped element configuration, a superconducting strip is properly shaped and sized in order to perform like a radiation absorber, and this structure, which is an inductor as well, is coupled to a capacitor to form a superconductive high quality factor resonator. In this way, the change in kinetic inductance, due to the incident radiation, produces a change in the resonant frequency and in the quality factor, which can be sensed by measuring the change in the amplitude and phase of the bias signal of the resonator, transmitted past the resonator through a feedline. 

The KID design and readout scheme are intrinsically multiplexable for large--format arrays, provided that the resonant frequencies of the individual resonators coupled to the same feedline are adjusted to unique values, for instance by changing the capacitor size. In this way, entire arrays can be fed and read out thanks to an electronics chain made of \emph{cold components}, including low noise amplifiers (LNAs), dc--blocks, coaxial cables, and power attenuators; and a \emph{room--temperature stage}, where an FPGA-based electronics, coupled to an ADC/DAC board, is used to generate one bias tone per resonator. This solution allows to feed and monitor the amplitude and phase of the bias signals of all the resonators at the same time, while physically connecting the cold stage to the room--temperature with one cable only. 

KID technology has been already proven in ground--based experiments \cite{Ritacco}, and for its features seems to be the optimal solution for next--generation space--borne CMB experiments \cite{1475-7516-2018-04-014,1475-7516-2018-04-015}, but it still needs to be demonstrated in a representative environment for space applications. OLIMPO, which was operated from the stratosphere, is therefore a natural testbed for KIDs in space--like conditions.

\section{Detectors and \emph{cold electronics}}

The first constraint in the optimization process of a detector system is always the target science for which it will be built. In the OLIMPO case, moreover, it has to fit an already developed cryogenic and optical system. This implies that the first step is the choice of the material of the superconducting film and the dielectric substrate, the size of the detector arrays, the geometry and size of the absorbers, the geometry and size of the radiation couplers, the number of detectors per array, and the illumination configuration. These steps have been performed through optical simulations. 

The second step concerns the optimization of the readout scheme: the geometry and size of the feedline; the geometry and size of the capacitors, on which the resonant frequencies of the resonators depend; and the coupling between the resonators and the feedline. This optimization has been done through electrical simulations.

The last step regards the optimization of the \emph{cold electronics}: the choice of the material and size of the coaxial cables; the magnitude of the power attenuators; the gain, noise, and operation temperature of the cryogenic amplifier.

\subsection{KID optimization, fabrication and results}

The detailed description of the optimization and fabrication of the OLIMPO detector systems and the measurement results can be found in \cite{Paiella2017,Paiella2018}. 

All the four arrays are fabricated in a \SI{30}{nm} thick aluminum film deposited on silicon substrates of different thickness depending on the observed radiation frequency. The substrate acts also as a backshort, since the face opposite to the detectors has been coated with aluminum. The properties of different aluminum film thicknesses have been measured as described in \cite{Paiella2016}, and the results have been used for the optical simulations. A compromise between optical simulation results and critical temperature (on which the optimal working temperature and the minimum detectable radiation frequency depend) forced us to choose \SI{30}{nm} for the aluminum film thickness. For this film, we measured the critical temperature, $T_{c}=\SI{1.31}{K}$; the residual resistance ratio, ${\rm RRR}=3.1$; the sheet resistance, $R_{s}=\SI{1.21}{\Omega/\Box}$, and the surface inductance, $L_{s}=\SI{1.38}{pH/\Box}$.

The optimal absorber solution results to be a front--illuminated IV order Hilbert pattern, where the characteristic length scales with the observed radiation wavelength. The 150 and 250 GHz arrays are coupled to the radiation via a single--mode waveguide, while the 350 and 460 GHz are coupled via a single--mode flared waveguide. The number of detectors per array is 23, 39, 25 and 43 for the 150, 250, 350 and \SI{460}{GHz} array, respectively.

The capacitors of the KIDs have been designed so that the lumped element condition is satisfied for all the resonators and the resonant frequencies fall into the range $\left[100;600\right]$ MHz. This is done by means of large capacitors, which have also the advantage of reducing the TLS (two--level system) noise \cite{Noroozian2009}. Moreover, the resonant frequencies are such that the \SI{150}{GHz} and the \SI{460}{GHz} arrays can be fed and read out with the same line as well as the \SI{250}{GHz} and the \SI{350}{GHz} arrays. In this way, each readout electronics manages about 65 detectors. Each detector is coupled to a $\SI{50}{\Omega}$--matched microstrip feedline (the width of which is different array by array) by means of capacitors, designed to constraint the coupling quality factor to about $1.5\times10^{4}$ guaranteeing, thus, a quite large detector dynamics, and that the total quality factor results dominated by the coupling one.

The arrays have been fabricated at the IFN--CNR. The detectors have been realized by electron--beam litography, electron--gun evaporation and lift--off \cite{Colantoni2016} on high--quality, high--resistivity ($\rho>\SI{10}{k\Omega.cm}$) intrinsic Si(100) wafers, double--side polished. The sample holders of the detector arrays are made of ergal alloy (aluminum 7075) as well as the horn arrays, in order to guarantee good thermalization and low power losses through the horns.

The OLIMPO detectors have been fully characterized: the electrical properties, such as the quality factors and the resonant frequencies, have been measured in a dark laboratory cryostat and result in agreement with the simulations; the optical performance has been measured in the OLIMPO cryostat and results in a Rayleigh-Jeans noise equivalent temperature lower than $\SI{500}{\mu K/\sqrt{Hz}}$ under a blackbody load of about \SI{90}{mK}, for all the OLIMPO arrays, and a global optical efficiency of about 10\%, averaged over the four channels.      

\subsection{Cold electronics optimization}

With \emph{cold electronics} we mean the components placed inside the cryostat, necessary to feed and read out the detectors. These components include coaxial cables, power attenuators, low noise amplifiers and dc--blocks. They have to be chosen in such a way that the noise equivalent power (NEP) of the active components of the \emph{cold electronics} is lower than the expected noise equivalent power of the detectors.

Theoretically, the noise of a KID is mainly due to the generation--recombination noise (the TLS noise can be neglected by design), thus its NEP can be evaluated as 
\begin{equation}
{\rm NEP}_{g-r}=2\Delta\sqrt{\frac{N_{qp}}{\tau_{qp}}}\;,
\end{equation}  
where $\Delta=1.764\;k_{B}T_{c}$ is half Cooper pair binding energy, with $k_{B}$ the Boltzmann constant, $N_{qp}$ is the quasiparticle number, and $\tau_{qp}$ is the quasiparticle lifetime (measured in \cite{Paiella2018}). For the OLIMPO detectors, the generation--recombination NEP is collected in table \ref{tab:NEP_gr}.

\begin{table}[h]
\caption{\small\label{tab:NEP_gr}Values of the generation--recombination NEP for the OLIMPO channels.}
\begin{center}
\begin{tabular}{cc}
\br
Channel&${\rm NEP}_{g-r}$\\
$\left[{\rm GHz}\right]$&$\left[{\rm W}/\sqrt{{\rm Hz}}\right]$\\
\mr
150&$3.1\times 10^{-17}$\\
250&$2.8\times 10^{-17}$\\
350&$2.1\times 10^{-17}$\\
460&$1.8\times 10^{-17}$\\
\br
\end{tabular}
\end{center}
\vspace{-0.5cm}
\end{table}  

As we already said, a KID is a superconductive resonator, which is optimally sensitive when operated close to its resonant frequency. This means that, since the quality factors are very high, due to the superconductor properties, and thus the resonances are very deep, the readout power has to be amplified. This is done by means of a cryogenic low noise amplifier, able to  amplify the signal output from the detectors, with very low intrinsic noise. For the OLIMPO experiment, the two LNAs, necessary for the two readout lines, have been provided by Arizona State University\footnote[1]{http://thz.asu.edu/products.html} (ASU) \cite{6881015}. These amplifiers dissipate about \SI{13}{mW} each ($V=\SI{1.6}{V}$ and $I=\SI{8}{mA}$ at \SI{10}{K}), and amplify \SI{33}{dB}, with a noise temperature of \SI{5}{K}, and a \SI{1}{dB} gain compression point, referred to the input, of \SI{-61}{dBm} at \SI{10}{K}, for a \SI{300}{MHz} signal. The noise equivalent power associated to the cryogenic amplifier is given by
\begin{equation}
{\rm NEP}_{amp}=\frac{N_{qp}\Delta}{\tau_{qp}}\sqrt{\frac{k_{B}T_{amp}}{P_{read}}}\;;
\label{eq:NEP_amp}
\end{equation}  
where $T_{amp}$ is the noise tempertaure, and $P_{read}$ is the total readout power at the LNA input.  

Since the OLIMPO cryostat does not feature a \SI{10}{K} stage, and since the coldest stage where they can be mounted is the vapor $^{4}$He shield, at a temperature of about \SI{30}{K}, we need to extrapolate the values of the LNA noise temperature at \SI{30}{K}, in order to compare the NEP associated to the LNA and the generation--recombination one. 

This has been done by combining the measurements provided by ASU at 10, 20 and \SI{300}{K}, with the measurements performed by us in a laboratory cryostat at 4 and \SI{44}{K}. Since \SI{44}{K} is the temperature closest to \SI{30}{K} at which we were able to cool the amplifier quickly, the measurements at this temperature have been performed at different supply voltages. All these data are collected in table \ref{tab:LNA}. The noise measurements have been done through an \emph{Anritsu MS2717B} spectrum analyzer, set to a resolution bandwith ${\rm RBW}=\SI{1}{Hz}$, at the LNA output, and have been scaled to the LNA input thanks to the gain measured with an \emph{Anritsu M52034B} vector network analyzer (VNA). All these measurements refer to \SI{300}{MHz}. The conversion between noise power and noise temperature is given by 
\begin{equation}
{\rm Noise\;Temperature}=\frac{{\rm Noise\;Power}}{k_{B}\:{\rm RBW}}\;.
\end{equation}  

\begin{table}[h]
\caption{\small\label{tab:LNA}Value of the supply voltage, gain, noise power and temperature at the input of the LNA provided by ASU, at different operation temperatures, at \SI{300}{MHz}.}
\begin{center}
		\begin{tabular}{cccccc}
		\br
		\multicolumn{1}{c}{\multirow{2}{*}{T}}&
		\multicolumn{1}{c}{\multirow{1}{*}{LNA}}&
		\multicolumn{1}{c}{\multirow{2}{*}{Gain}}&
		\multicolumn{1}{c}{\multirow{1}{*}{Noise Power}}&
		\multicolumn{1}{c}{\multirow{1}{*}{Noise Temperature}}&
		\multicolumn{1}{c}{\multirow{3}{*}{Notes}}\\
		
		\multicolumn{1}{c}{\multirow{1}{*}{}}&
		\multicolumn{1}{c}{\multirow{1}{*}{Voltage}}&
		\multicolumn{1}{c}{\multirow{1}{*}{}}&
		\multicolumn{1}{c}{\multirow{1}{*}{@ LNA input}}&
		\multicolumn{1}{c}{\multirow{1}{*}{$T_{amp}$}}&
		\multicolumn{1}{c}{\multirow{1}{*}{}}\\
		
		\multicolumn{1}{c}{\multirow{1}{*}{$\left[{\rm K}\right]$}}&
		\multicolumn{1}{c}{\multirow{1}{*}{ $\left[{\rm V}\right]$}}&
		\multicolumn{1}{c}{\multirow{1}{*}{$\left[{\rm dB}\right]$}}&
		\multicolumn{1}{c}{\multirow{1}{*}{$\left[{\rm dBm}\right]$}}&
		\multicolumn{1}{c}{\multirow{1}{*}{$\left[{\rm K}\right]$}}&
		\multicolumn{1}{c}{}\\
\mr
		4 & 1.6 & 33  & $-$192.0$\pm$0.2  &4.57$\pm$0.21&\\	
		44 & 1.5 & 32.1  & $-$189.1$\pm$0.2  &8.91$\pm$0.41&\\
		44 & 1.6 & 32.4  &  $-$189.4$\pm$0.2&8.31$\pm$0.38&\\
		44 & 1.7 & 32.7  &  $-$189.7$\pm$0.2&7.76$\pm$0.36&Measurements performed \\
		44 & 1.8 & 32.9 &   $-$189.9$\pm$0.2&7.41$\pm$0.34&by us in a laboratory cryostat\\
		44 & 1.9  & 33.0& $-$190.0$\pm$0.2&7.24$\pm$0.33&\\
		44 & 2.0 & 33.1  & $-$190.1$\pm$0.2&7.08$\pm$0.33&\\
		44 & 2.1 & 33.2 &  $-$190.2$\pm$0.2&6.92$\pm$0.32&\\
		\mr
		10 & 1.6 & 33 &  & 5&\\
		20 & 1.6 &  &  & 6&provided by ASU\\
		300 & 2.1 &30&  & 45&\\
\br
\end{tabular}
\end{center}
\end{table}  

\begin{figure}[h]
\begin{center}
\includegraphics[scale=0.40]{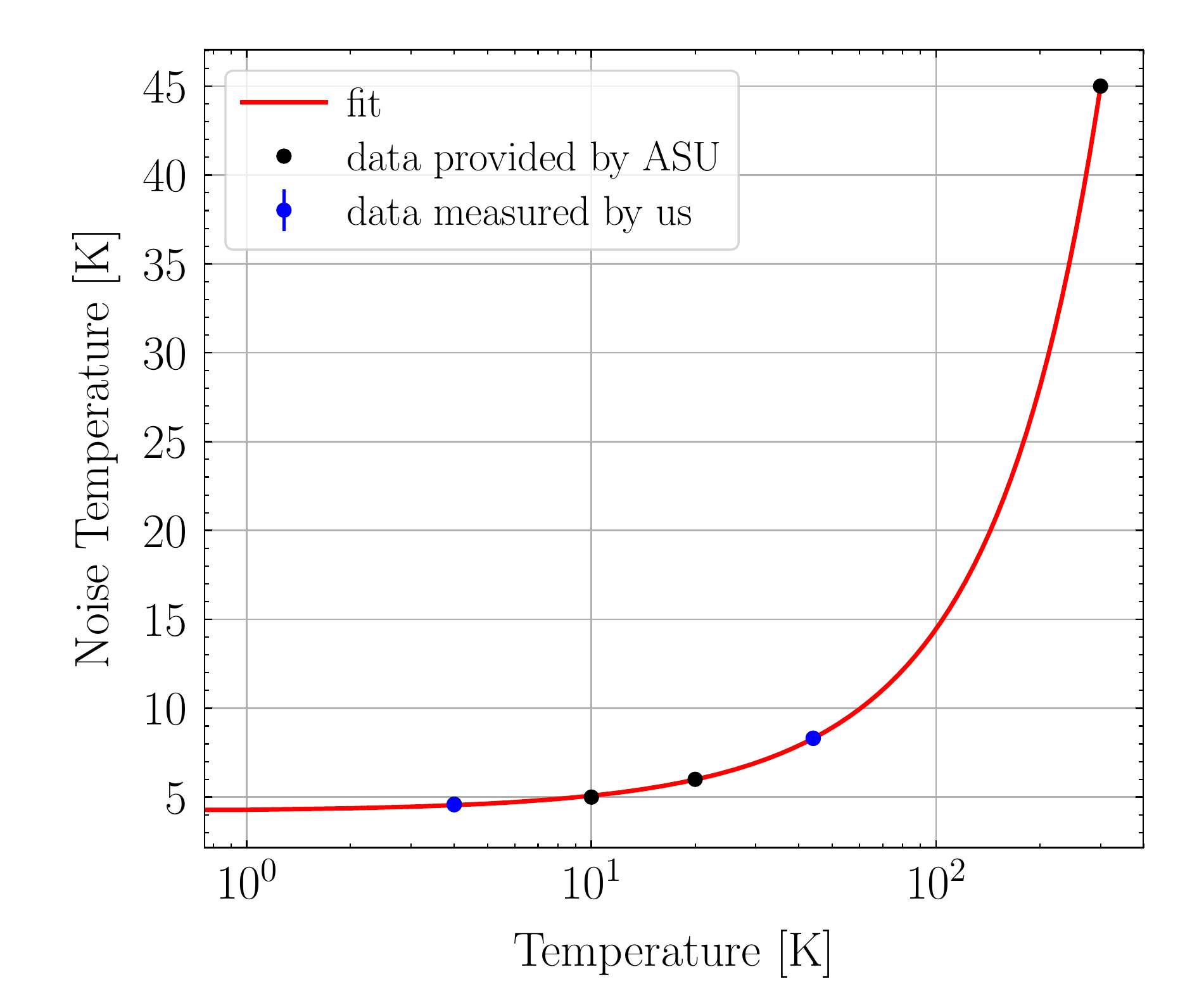}
\includegraphics[scale=0.40]{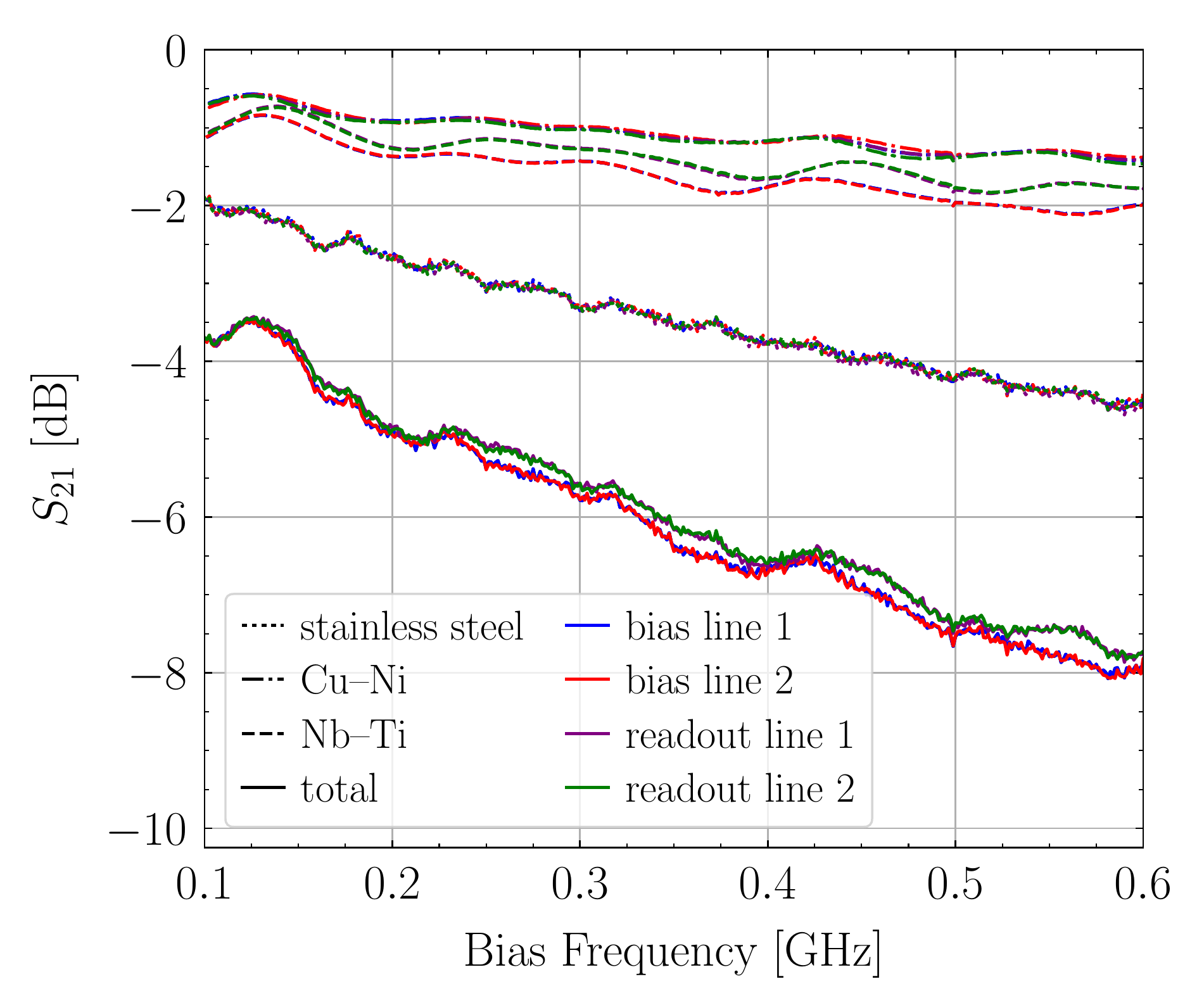}
\end{center} 
\caption{\small\label{fig:LNA} \emph{Left panel}: Measured data (\emph{black} and \emph{blue} dots) and fit (\emph{red solid line}) of the noise temperature at the LNA input as function of the operation temperature. \emph{Right panel}: $S_{21}$ parameter measured for the coaxial cables at room--temperature. \emph{Dotted lines} are for stainless steel cables, \emph{dot--dash lines} are for Cu--Ni cables, \emph{dashed lines} are for Nb--Ti cables, and \emph{solid lines} represent the whole lines. Different colors indicate the different lines where the cables are.}
\end{figure}

The value of the noise temperature of the LNA at \SI{30}{K} has been extrapolated by fitting the data with the function $aT^{2}+bT+c$, see the \emph{left panel} of figure \ref{fig:LNA}. We obtained $a=\SI{1.67e-4}{K^{-1}}$, $b=8.59\times10^{-2}$, and $c=\SI{4.20}{K}$, and therefore

\begin{equation}
T_{amp}\left(\SI{30}{K}\right)=\SI{6.93}{K}\;.
\end{equation}  

The only information which is missing to calculate the LNA NEP via equation \ref{eq:NEP_amp} is $P_{read}$. The total readout power at the LNA input is the sum of all the bias powers of the tones, each attenuated by the corresponding resonator, on the same readout line. Moreover, in order to work in the linear regime of all the amplifier circuits of the \emph{cold} and \emph{room--temperature electronics}, the powers at the input of such amplifiers have to be lower than the 1 dB gain compression point of the amplifiers themselves. As we are going to see in section \ref{sec:room-temperature}, the amplifier room--temperature components have been chosen and ordered along the readout chain in such a way that the 1 dB gain compression points are matched, and therefore the only constraint on $P_{read}$ is given by the first amplifier: the LNA.

OLIMPO KIDs have been designed to be fed with a bias power of about \SI{-80}{dBm} each, and with a resonance deep of about \SI{10}{dB}. This means that, considering an average of 65 detectors per readout line, the total readout power at the LNA input is $P_{read}=\SI{-72}{dBm}$, below the \SI{1}{dB} gain compression point of \SI{-61}{dBm}. Therefore, using equation \ref{eq:NEP_amp}, we evaluated the LNA NEPs, collected in table \ref{tab:NEP_amp}, which result lower than the generation--recombination ones of table \ref{tab:NEP_gr}. The two LNAs have thus been safely mounted on the \SI{30}{K} shield of the OLIMPO cryostat.

 \begin{table}[h]
\caption{\small\label{tab:NEP_amp}Values of the LNA NEP for the OLIMPO channels.}
\begin{center}
\begin{tabular}{cc}
\br
Channel&${\rm NEP}_{amp}$\\
$\left[{\rm GHz}\right]$&$\left[{\rm W}/\sqrt{{\rm Hz}}\right]$\\
\mr
150&$9.3\times 10^{-18}$\\
250&$7.6\times 10^{-18}$\\
350&$4.1\times 10^{-18}$\\
460&$3.0\times 10^{-18}$\\
\br
\end{tabular}
\end{center}
\vspace{-0.5cm}
\end{table}  

The cryogenic coaxial cables complete the \emph{cold electronics}. We equipped the OLIMPO cryostat with coaxial cables made of three different materials, for the different thermal jumps: stainless steel from \SI{300}{K} to \SI{30}{K} and \emph{viceversa}, Cu--Ni from \SI{30}{K} to \SI{1.8}{K} and \emph{viceversa}, and Nb--Ti from \SI{1.8}{K} to \SI{0.3}{K} and \emph{viceversa}. In particular, the bias line from \SI{300}{K} to \SI{0.3}{K} is composed of a \SI{2}{m} long stainless steel cable, a \SI{350}{mm} long Cu--Ni cable, and a \SI{400}{mm} long Nb--Ti cable; the readout line from \SI{0.3}{K} to \SI{300}{K} is composed of a \SI{350}{mm} long Nb--Ti cable, a \SI{350}{mm} long Cu--Ni cable, and a \SI{2}{m} long stainless steel cable. The power losses, in terms of $S_{21}$ parameter, have been measured with the VNA at room--temperature and are shown in the \emph{right panel} of figure \ref{fig:LNA}. The maximum power loss of the bias line, in the frequency range of interest, at room--temperature is about 8 dB, which surely decreases at cryogenic temperature. 

Along the bias line, we have inserted three cryogenic power attenuators of magnitude \SI{10}{dB} each, thermalized at \SI{1.8}{K} so that their contribution to the total noise results negligible. This is necessary to guarantee a large dynamics for the bias power sweeps on the resonators, essential to find the optimal working point; remaining in the linear regime of the LNA. Precisely, remembering that the LNA \SI{1}{dB} gain compression point is \SI{-61}{dBm}, which means \SI{-69}{dBm} per tone at the detector array input, and since our \emph{room--temperature readout electronics} can send a maximum total power of \SI{-3}{dBm}, which means \SI{-21}{dBm} per tone, we need to attenuate \SI{48}{dB}. In first approximation, we have seen that the bias line attenuates at most \SI{8}{dB}, and the bias room--temperature coaxial cable (between the \emph{room--temperature readout electronics} and the cryostat) attenuates 8 dB, therefore we have to attenuate at least \SI{32}{dB}.

\section{\label{sec:room-temperature}\emph{Room--temperature readout electronics}} 

\begin{figure}[h]
\begin{center}
\includegraphics[scale=0.60]{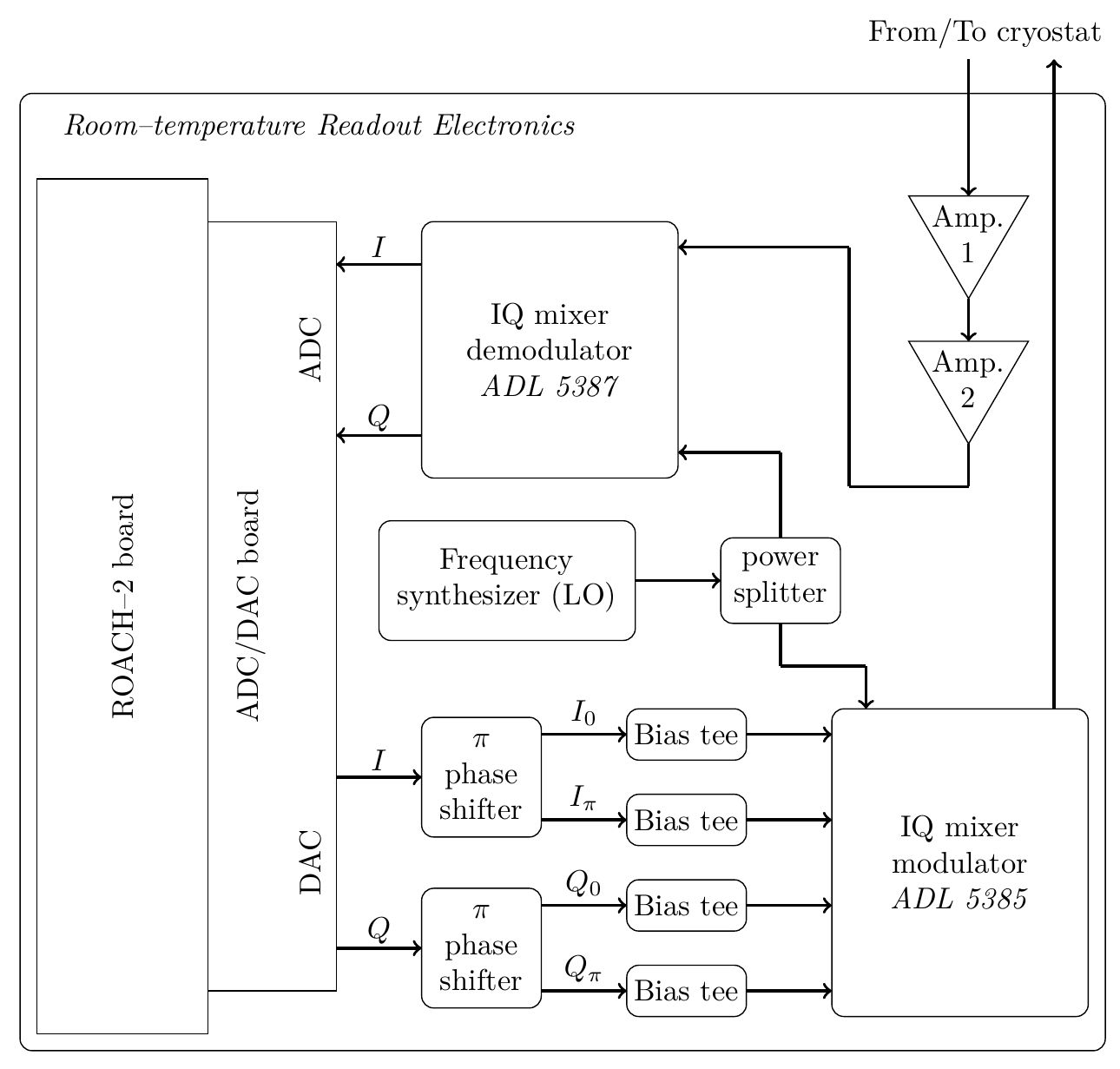}
\includegraphics[scale=0.41]{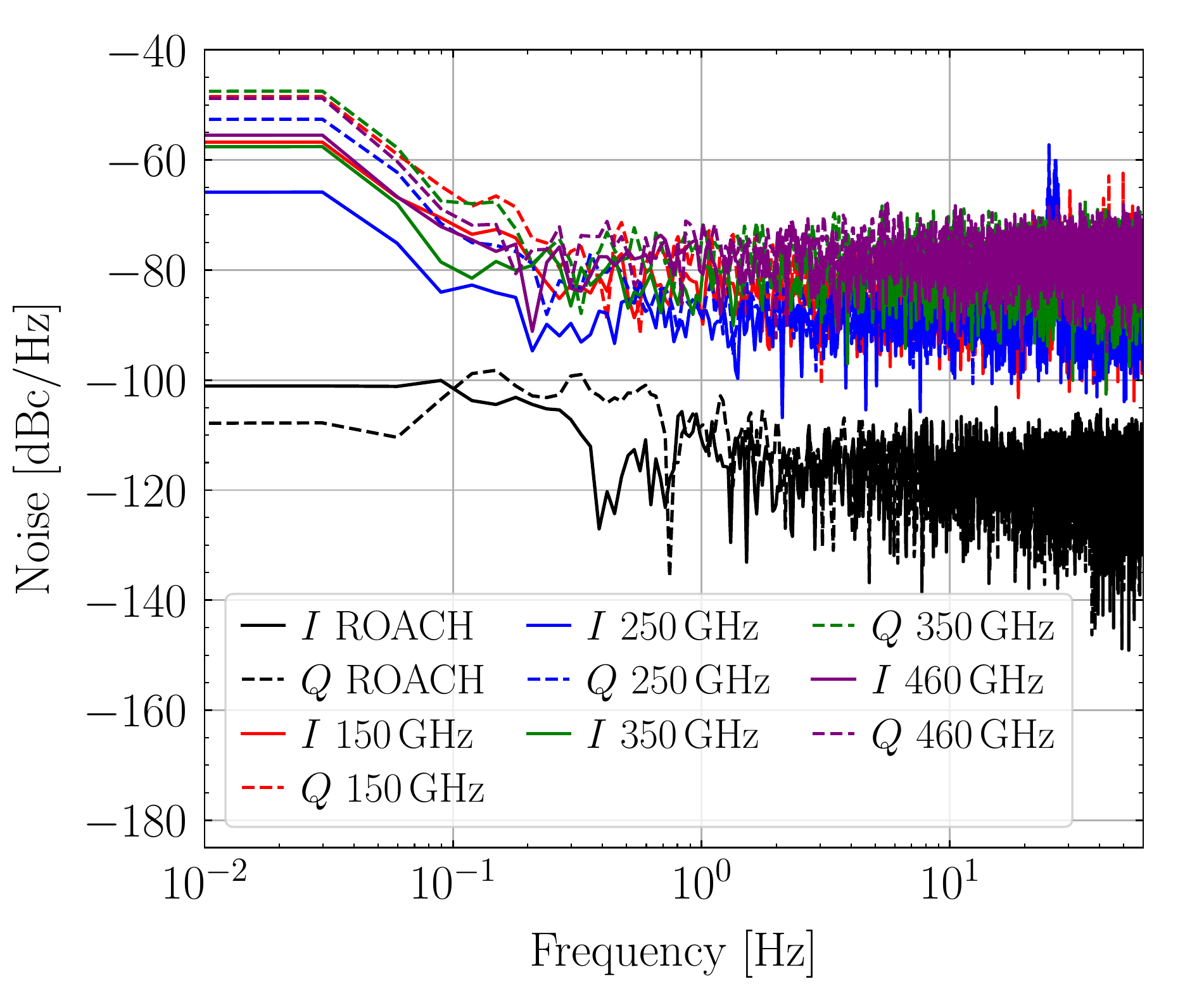}
\end{center} 
\caption{\small\label{fig:roach} \emph{Left panel}: Block diagram of the \emph{room--temperature electronics}. The ROACH--2  and MUSIC DAC/ADC boards are completed by a set of microwave components to form the bias--readout electronics. \emph{Right panel}: Measured noises of the \emph{room--temperature electronics} (\emph{black lines}) closed on a \SI{36}{dB} attenuator (loopback mode) and of the OLIMPO detector arrays (\emph{color lines}), for both the $I$ (\emph{solid lines}) and $Q$ (\emph{dashed lines}) channels.}
\end{figure}

The detection system is completed by the \emph{room--temperature electronics}. Our FPGA consists of a ROACH--2 board\footnote{https://casper.berkeley.edu/wiki/ROACH2}, coupled to a MUSIC DAC/ADC board\footnote{https://casper.berkeley.edu/wiki/MUSIC\_Readout}, the firmware of which has been developed by ASU, and is able to generate up to 1000 tones over a \SI{512}{MHz} bandwidth, with a demodulated output sampling rate up to about \SI{1}{kHz} \cite{Gordon2016}. Since the resonant frequencies of the OLIMPO resonators include values higher than \SI{256}{MHz}, the electronics has to be equipped with up--conversion and down--conversion microwave components. The block diagram of such electronics is shown in the \emph{left panel} of figure \ref{fig:roach}.

The microwave components have been chosen in such a way to optimize the bias--readout electronics in terms of noise, bias and readout power, and to work in the linear regime of the amplifier components (amplifiers and demodulator). The IQ mixer modulator requires as input the $I$ and $Q$ signals and their $\pi$--phase shifts, offset positive through four bias tees. The maximum total power delivered at the IQ modulator output is \SI{-3}{dBm}, and the noise floor is \SI{-159}{dBm/Hz}. The room--temperature amplifiers have been selected and located along the readout line in such a way that the power budget allow them and the IQ demodulator to work in the linear regime (the power at the input of such components is lower than the 1 dB gain compression point at the input) and that the total noise figure at the demodulator output was as low as possible. Table \ref{tab:warm_amp} shows the specifications of the two room--temperature amplifiers and the IQ demodulator, which result in a total gain ${\rm G}=\SI{46.9}{dB}$ and a total noise figure ${\rm NF}=\SI{0.44}{dB}$, both estimated at the demodulator output. The NF has been calculated as
\begin{equation}
{\rm NF}=10\log_{10}\left(n_{1}+\sum_{i=2}^{3}\frac{n_{i}-1}{\prod_{j=1}^{i-1}g_{j}}\right)\;,
\end{equation}
 where $n_{i}=10^{{\rm NF}_{i}/10}$, $g_{j}=10^{{\rm G}_{j}/10}$, $i,j=1$ is Amp. 1, $i,j=2$ is Amp. 2, and $i,j=3$ is the IQ demodulator.

 \begin{table}[h]
\caption{\small\label{tab:warm_amp}Specifications of the amplifiers components.}
\begin{center}
\begin{tabular}{ccccc}
\br
\multirow{2}{*}{Component}&Gain (G)&\multicolumn{2}{c}{\SI{1}{dB} gain compression point}&Noise Figure (NF)\\
&$\left[{\rm dB}\right]$&@ input $\left[{\rm dBm}\right]$& @ output $\left[{\rm dBm}\right]$&$\left[{\rm dB}\right]$\\
\mr
Amp. 1&20&2&22&0.4\\
Amp. 2&22.5&$-$9.5&13&2.7\\
IQ demodulator&4.4&13&17.4&13.2\\
\br
\end{tabular}
\end{center}
\vspace{-0.5cm}
\end{table}  

For this electronics, we measured the loopback noise by closing it on a power attenuator of \SI{36}{dB}, in such a way to have \SI{-56}{dBm} at the Amp. 1 input (total power of \SI{-20}{dBm} at the output of the IQ modulator), similar to that we expect at the output of the LNA (\SI{-40}{dBm}), attenuated by the cryogenic and room--temperature readout line (\SI{-16}{dB}). In this way, the noise floor of the IQ modulator is reduced to the room--temperature thermal noise: \SI{-174}{dBm/Hz}. The measured loopback noise, shown in the \emph{right panel} of figure \ref{fig:roach} compared to the OLIMPO detector array noises, results $\left(-117\pm1\right)\SI{}{dBc/Hz}$, which means $\left(-126.1\pm1.0\right)\SI{}{dBm/Hz}$, for both the $I$ and the $Q$ channels. The expected noise can be estimated as
\begin{equation}
{\rm Noise}=\SI{-174}{dBm/Hz}+{\rm G}+{\rm NF}=\SI{-126.6}{dBm/Hz}\;,
\end{equation}   
which is compatible with the measured one. Therefore the microwave components have been correctly optimized, and the \emph{readout electronics} noise is indeed lower than that measured on the four OLIMPO detector arrays, as shown in the \emph{right panel} of figure \ref{fig:roach}.

\section{Conclusion}
We have designed, optimized, fabricated and characterized four arrays of horn--coupled LEKIDs, able to work in the OLIMPO experiment. In this paper we have focused the attention on the optimization process of the \emph{cold} and \emph{room--temperature electronics}.

The \emph{cold electronics} has been optimized in terms of gain, noise, and operation temperature of the LNA; power attenuation and temperature thermalization of the coaxial wiring; and magnitude of the power attenuators. The \emph{room--temperature electronics} has been optimized in such a way that the generated tones were at the resonant frequencies of the KIDs; the signal powers were in the dynamics of each microwave components; and the loopback noise was lower than the one measured for the detector arrays.   

All the work has been done by extrapolating the LNA noise at 30 K from measurements at different temperatures; and by measuring the $S_{21}$ parameter of the coaxial cables, the loopback noise of the \emph{room--temperature electronics}, and the noise of the KID arrays. 

We obtained that the LNA noise is lower than the expected KID noise as well as the loopback noise of the \emph{room--temperature electronics} is lower than the measured KID noise. 

\bibliography{bib_abbr}

\end{document}